**Comment on the history of the *stretched exponential function***

*M. Cardona*[1,*]*, R.V. Chamberlin*[2]*, and W. Marx*[1]

[1] *Max Planck Institute for Solid State Research, Heisenbergstr. 1, D-70569 Stuttgart, Germany*
[2] *Department of Physics, Arizona State University, Tempe, AZ 85287-1504, USA*

* Corresponding author E-mail: m.cardona@fkf.mpg.de

The current citations of 19th century papers have recently received considerable attention [1]. Among them is an article by Rudolf (Hermann Arndt) Kohlrausch (*1809, [†]1858) [2] in which he introduced the stretched exponential function $Q_t = Q_0 [\exp(-Bt^\alpha)]$ to interpret charge relaxation in a Leiden jar. This work was essentially forgotten throughout most of the 20th century until the appearance of Ref. 3. Unfortunately, R.G. Palmer et al. [3] do not refer to the correct paper by R. Kohlrausch [2], referring instead to an earlier Kohlrausch paper [4] in which he only qualitatively describes relaxation phenomena in galvanometer threads. The fact that such mechanical relaxation can also be represented by the stretched exponential function was not published until many years later by Friedrich (Wilhelm Georg) Kohlrausch (*1840, [†]1910) [5], the son of Rudolf Kohlrausch. Therefore, the credit for the introduction of the concept is sometimes given to him. But the first time the stretched exponential function appeared in the literature is 1854 in [2].

The stretched exponential function was rediscovered in 1970 by Graham Williams and David C. Watts [6], who independently developed an equivalent expression and re-introduced the concept, stimulating its application to relaxation and scattering phenomena together with theoretical studies. Accordingly, the function was termed the Williams-Watts (WW) function or the Kohlrausch-Williams-Watts (KWW) function. The Williams & Watts paper [6] achieved an exceptionally high impact, currently with 2143 citations. We note that a complementary expression known as the Weibull distribution [7] evolved a similar impact, primarily in the materials science and engineering literature. To complete the story: The term "stretched exponential" is first found in [8]. Since 1900 Ref. 2 was first cited in 1906 by A. Joffe in [9] and Ref. 3 was first cited in 1943 by H. Leaderman in [10].

Ever since the highly cited Palmer paper [3] (1019 citations) appeared most papers referring to Kohlrausch's work give the wrong reference [4]. This is clearly demonstrated in Fig. 1 where the citations per year to [4] are shown to rise sharply after the appearance of [3]. Additional proof that Ref. 3 is the source of the error propagation is obtained by looking at how many papers cite [3] and [4] simultaneously (co-citations). The number of co-citations of these two papers is 143 (the first one appearing in 1985) whereas the number of those co-citing [2] and [3] is only 43. The Palmer paper [3], however, was not the first one to cite Ref. 4 instead of Ref. 2. There are three pre-1984 citing papers which cited Ref. 4, but they received altogether less than 50 citations and are therefore unlikely to be responsible for the propagation of the confusion.

Initially, Williams & Watts were not aware of the early Kohlrausch papers and hence did not refer to [2] or [4]. But their paper obviously was a strong stimulus for the subsequent widespread application of the concept of the stretched exponential function and thereby caused the revival of the early works of R. Kohlrausch within the scientific literature. Co-citation analysis of the Williams & Watts paper [6] reveals that (1) almost 70 percent (716 out of 1052) of the total Kohlrausch citations were co-cited with [6], and that (2) the number of co-citations of [4] and [6] is still higher (475) than of [2] and [6] (257). Ref. 2 and Ref. 4 are both not at all co-cited with the Weibull paper [7]. A summary of the citation data is shown in Table 1.

For comparison we show in Fig. 2 the citation history (citations per year) of [3], of [6], and of [7], together with the sum of [2] and [4]. Notice that while the citations of the papers by Williams & Watts [6] and by Weibull [7] increase with time, the citations of the Palmer paper [3] have been continuously decreasing since 1987. The sum of the citations of both Kohlrausch papers oscillates around 60 citations per year during the past 15 years. The yearly rate of incorrect citations far exceeded the rate of correct citations until the year 2002, indicating that the citation error has been slowly realized. In the following year, however, the time-curves intersected again, so that the total number of incorrect citations (713 since 1900, including 21 in PRL) remains much higher than the total number of correct citations (368 since 1900, including 5 in PRL). Although several papers (e.g. Ref. 11) mentioned the incorrectness in the recent past, a reversal of the trend seems to be unlikely.

For the sake of completeness we note that many more papers (757 since 1991) have mentioned explicitly the name Kohlrausch (164 together with the term "stretched exponential"), 593 without citing any original source. This reference phenomenon is referred to as "informal citation" [12]. The citation data mentioned here are based on the Thomson/ISI Web of Science (WoS) covering the time period 1900 till present (currently, the pre-1900 citations of papers published before 1900 are not available).

Although the name Kohlrausch is not very common, we have investigated the possibility of additional errors related to homonyms of Kohlrausch. To start with, a few words concerning Friedrich Wilhelm Kohlrausch [5]: Since 1900 we find altogether 2019 citing papers referring to publications by F. W. Kohlrausch. Most of them (573) relate to his book "Praktische Physik" which first appeared in 1870 and has had a large number of editions since; the last one appeared in 1996 [13]. We also find articles for additional 12 Kohlrauschs, most of them lowly cited, except for Karl Wilhelm Friedrich Kohlrausch (*1884, †1953), an Austrian (grandson of Rudolf Kohlrausch and nephew of Friedrich Kohlrausch) who published 77 articles in journals and many more in books and reports, mostly (59) about the Raman effect.

K.W.F. Kohlrausch is also responsible for some confusion in the literature concerning the Raman effect: Some of his early articles, the first one published 1931 [14], and a book [15] on this subject, refer to it as the Smekal-Raman effect. A. Smekal was a colleague of K.W.F. Kohlrausch who, five years prior to Raman's experimental discovery in 1928, wrote in 1923 a short article [16] on the theory of inelastic scattering (114 citations). Fortunately, K.W.F. Kohlrausch soon gave up his attempt to establish Smekal as a co-discoverer of the Raman effect. Although the term Smekal-Raman is found almost exclusively in the Austrian and the German literature, it first appeared in an American journal [17].

| Papers | # Citations since 1900 | Year of first Citation since 1900 | # Co-Citations since 1900 |
|---|---|---|---|
| Ref. 2 | 368 | 1906 | |
| Ref. 4 | 713 | 1943 | |
| Ref. 5 | 177 | 1910 | |
| Ref. 2 OR Ref. 4 | 1052 | 1906 | |
| Ref. 2 AND Ref. 4 | | 1977 | 29 |
| Ref. 3 | 1019 | 1985 | |
| Ref. 2 AND Ref. 3 | | 1986 | 43 |
| Ref. 4 AND Ref. 3 | | 1985 | 143 |
| Ref. 6 | 2143 | 1970 | |
| Ref. 7 | 2387 | 1952 | |
| Ref. 2 AND Ref. 6 | | 1984 | 257 |
| Ref. 4 AND Ref. 6 | | 1983 | 475 |
| (Ref. 2 OR Ref. 4) AND Ref. 6 | | 1984 | 716 |

**Table 1:** Summary of the citation data of Ref. 2 to Ref. 7, including the co-citations of Ref. 2 as well as Ref. 4 along with Ref. 3 and Ref. 6 (date of searching: 2007-09-06).

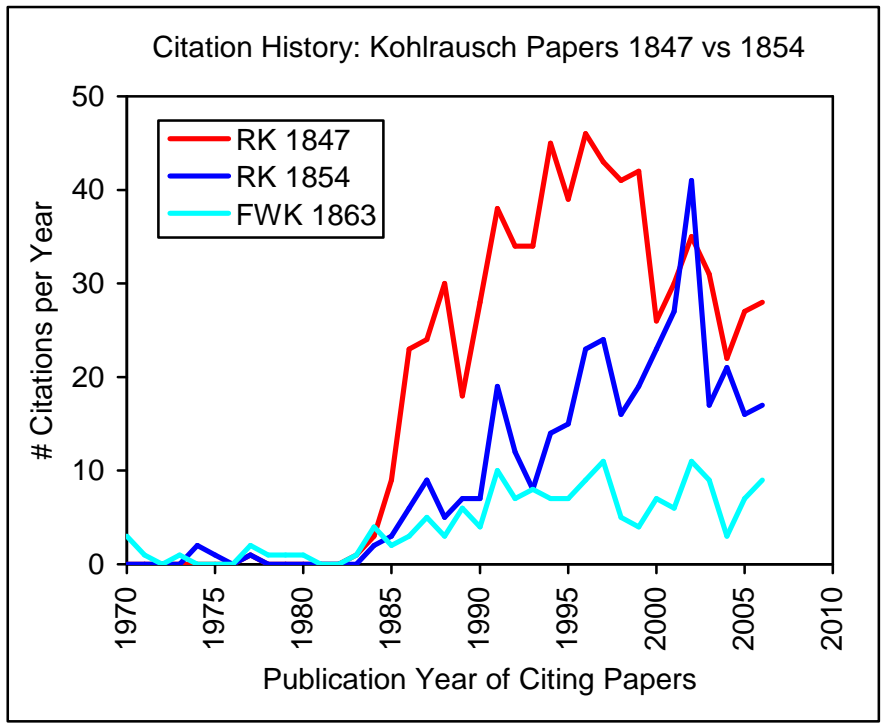

**Figure 1**: Citation history of Ref. 2, of Ref. 4, and of Ref. 5 (RK = Rudolf Kohlrausch, FWK = Friedrich Wilhelm Kohlrausch).

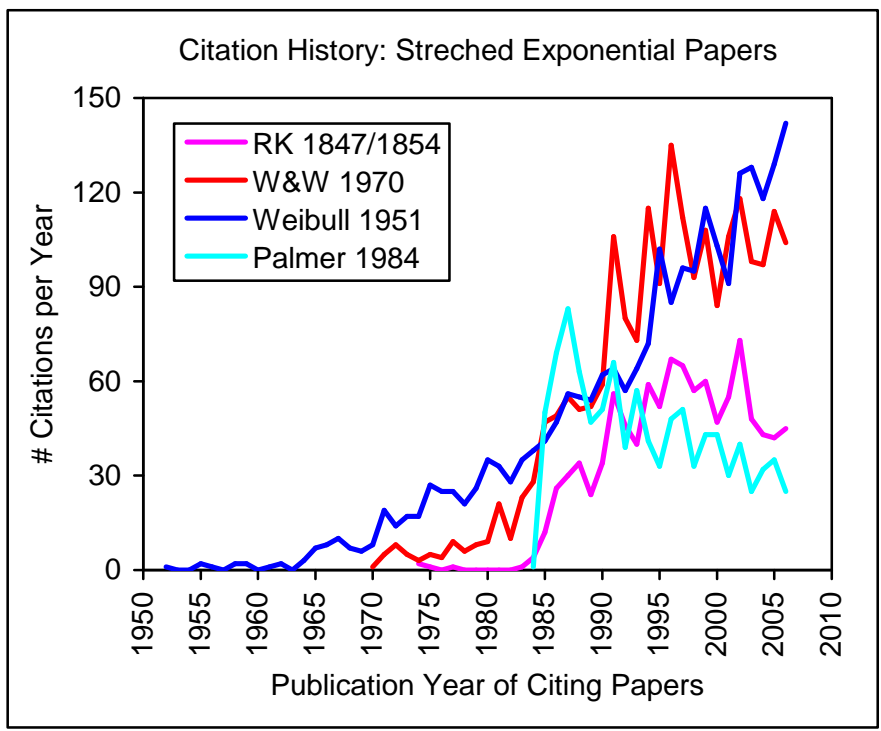

**Figure 2**: Citation history of the sum of Ref. 2 and Ref. 4, of Ref. 3, of Ref. 6, and of Ref. 7 (RK = Rudolf Kohlrausch, W&W = Williams & Watts).